\documentclass
[floatfix,superscriptaddress,secnumarabic,amssymb,amsmath,nobibnotes,aps,prd,showkeys,showpacs,nofootinbib,onecolumn,notitlepage,12pt]{revtex4}%
\usepackage{setspace}
\usepackage{color}
\usepackage{amsmath}
\usepackage{amsfonts}
\usepackage{verbatim}
\usepackage{amssymb}
\usepackage{graphicx,bm}
\usepackage{graphicx}
\usepackage{amsmath}
\usepackage{amssymb}
\usepackage{amssymb}
\usepackage{graphicx,bm}
\usepackage{graphicx}
\usepackage[caption=false]{subfig}%
\providecommand{\U}[1]{\protect\rule{.1in}{.1in}}
\newcommand{\be}{\begin{equation}}
\newcommand{\ee}{\end{equation}}

\newcommand{\mincir}{\raise
-3.truept\hbox{\rlap{\hbox{$\sim$}}\raise4.truept\hbox{$<$}\ }}
\newcommand{\magcir}{\raise
-3.truept\hbox{\rlap{\hbox{$\sim$}}\raise4.truept\hbox{$>$}\ }}

\ifx\pdfoutput\relax\let\pdfoutput=\undefined\fi
\newcount\msipdfoutput
\ifx\pdfoutput\undefined\else
\ifcase\pdfoutput\else
\ifx\paperwidth\undefined\else
\ifdim\paperheight=0pt\relax\else\pdfpageheight\paperheight\fi
\ifdim\paperwidth=0pt\relax\else\pdfpagewidth\paperwidth\fi
\fi\fi\fi
\begin{document}
\title{The Common Solution Space of General Relativity}
\author{Andronikos Paliathanasis}
\email{anpaliat@phys.uoa.gr}
\affiliation{Institute of Systems Science, Durban University of Technology, Durban 4000,
South Africa}
\affiliation{Departamento de Matem\'{a}ticas, Universidad Cat\'{o}lica del Norte, Avda.
Angamos 0610, Casilla 1280 Antofagasta, Chile}

\begin{abstract}
We review the solution space for the field equations of Einstein's General
Relativity for various static, spherically symmetric spacetimes. We consider
the vacuum case, represented by the Schwarzschild black hole; the de
Sitter-Schwarzschild geometry, which includes a cosmological constant; the
Reissner-Nordstr\"{o}m geometry, which accounts for the presence of charge.
Additionally we consider the homogenenous and anisotropic locally rotational
Bianchi II spacetime in the vacuum. Our analysis reveals that the field
equations for these scenarios share a common three-dimensional group of point
transformations, with the generators being the elements of the $D\otimes
_{s}T_{2}$ Lie algebra, known as the semidirect product of dilations and
translations in the plane. Due to this algebraic property the field equations
for the aforementioned gravitational models can be expressed in the equivalent
form of the null geodesic equations for conformally flat geometries.
Consequently, the solution space for the field equations is common, and it is
the solution space for the free particle in a flat space. This appoach open
new directions on the construction of analytic solutions in gravitational
physics and cosmology.

\end{abstract}
\keywords{Exact solutions; General Relativity; Solution Space; Blach holes; minisuperspace}\maketitle

\section{Introduction}

\label{sec1}

The construction of exact and analytic solutions for differential equations is
essential in all areas of physics and applied mathematics. Closed-form
solutions provide critical insights into the dynamical behavior of the given
system and enhance our understanding of the initial value problem. The
importancy of the analytic solutions is well described by Arscott in the
introduction of his book \cite{asc}.

Nowadays because of the high computing capacity allows for the numerical
treatment of nonlinear differential equations. While numerical methods offer
immediate information about the local behavior of the dynamical system near
the initial conditions, they do not always provide a global understanding of
the solution's behavior and information regarding the initial value problem
\cite{ch1}.

A powerful approach for the analytic treatment of nonlinear differential
equations is the symmetry analysis established by Sophus Lie
\cite{lie1,lie2,lie3}. The key characteristic of symmetry analysis is the
identification of invariance properties of differential equations under finite
transformations related to continuous groups. The presence of a symmetry
vector indicates the existence of invariant functions, which can be used to
simplify the given differential equation by defining a new reduced equation.
When feasible, this approach allows for the construction of closed-form
solutions, known as similarity solutions, for more details we refer the reader
to \cite{olver,kumei,ibra,Ovsi}.\ 

A pioneer application of the symmetry analysis established by Emmy Noether
\cite{Noether18}. Criteria have been established where the admitted symmetries
are directly related to the existence of conservation laws for a given
dynamical system. While conservation laws can be derived through various
approaches other than Noether's theorems \cite{hj1,hj2}, the simplicity and
systematic nature of Noether's algorithm make her work one of the most
influential studies in physical science \cite{noe1,noe2,noe3,noe4,noe5,noe8}.

In gravitational physics, symmetries play a crucial role at every stage of the
theory. The general form of the physical space is constrained by the existence
of symmetries \cite{sym1,sym2,sym3,sym4,sym5}. Moreover, the differential
equations governing the dynamical variables, as they are provided by General
Relativity, are nonlinear second-order differential equations. The symmetry
method has been widely applied to these equations to determine solutions see
for instance \cite{sym6,sym7,sym8,sym9,sym10,sym11} and references therein.

In this study, we review the solution space for the field equations of several
well-known gravitational models in General Relativity. We explore the
dynamical behavior of Einstein's field equations and identify common features
within the solution space for these spacetimes. This study extends the concept
of symmetries in gravitational models and opens new directions for
constructing analytic solutions. We demonstrate that the field equations for
the gravitational models under consideration can be solved using the
linearization approach \cite{ln1,ln2}. Specifically, we show that these field
equations can be reformulated as a set of linear equations, rendering the
dynamics trivial. Furthermore, we illustrate that the solutions to Einstein's
field equations can be expressed in terms of linear functions. The structure
of the paper is as follows.

In Section \ref{sec2} we present the basic properties and definitions for the
conformally related metrics and Lagrangians. In the following Sections we
investigate the solution space for the Einstein's field equations in various
models. In Section \ref{sec3}, we study the solution space for the field
equations of the static spherical symmetric spacetime in the vacuum leading to
the Schwarzschild black hole.\ Furthermore, in Section \ref{sec4} the
cosmological constant is introduced, where we show that the de
Sitter-Schwarzschild solution it has the same origin with that of the vacuum
spacetime. In \ref{sec5} we introduce charge and we study the solution space
for the Reissner-Nordstr\"{o}m black hole. We extend our analysis to the
cosmological case and specifically to the homogeneous and anisotropic locally
rotational Bianchi II spacetime. The solution space for this gravitational
model is determined in Section \ref{sec6}.

For all these gravitational models, the field equations can be linearized
through point transformations, which means that they share the solution space.
Specifically, the Einstein's field equations can be written in the equivalent
form of the geodesic equations in a flat space. The origin for this common
feature is discussed in Section \ref{sec7}. Finally, in Section \ref{sec8} we
summarize our results and we draw our conclusions.

\section{Conformally related metrics and Lagrangians}

\label{sec2}

In this Section we briefly discuss the basic mathematical definitions
necessary for the rest of the study.

\subsection{Conformally related metrics}

Consider the two metric tensor $g_{ij},~\bar{g}_{ij}.$ We say that the tensors
$g_{ij},~\bar{g}_{ij}$ are conformally related if there exist a function
$\Omega\left(  x^{k}\right)  $, such that, $\bar{g}_{ij}=\Omega^{2}g_{ij}%
~$\cite{sym4}.

As conformal symmetries are characterized the generators $X$ of the point
transformations which preserves the angles between two lines and the null
structure.\ In the case where not only angles and the null structure but also
the length is preserved, the CKV will be characterized as a killing symmetry
(KV) or isometry.

Let $X$ be a CKV for the metric tensor $g_{ij}$, the following mathematical
condition holds true \cite{sym4}
\begin{equation}
\mathcal{L}_{X}g_{ij}=2\psi\left(  x^{k}\right)  g_{ij}~,~
\end{equation}
\bigskip where $\psi\left(  x^{k}\right)  $ is known as the conformal factor
defined as $\psi\left(  x^{k}\right)  =\frac{1}{\dim g}X_{~;k}^{k}$, and
$\mathcal{L}_{X}$ is the Lie derivative with respect to the vector field $X$.

For the conformally related metric $\bar{g}_{ij}$ the symmetry condition for
the CKV reads \cite{sym4}
\begin{equation}
\mathcal{L}_{X}\bar{g}_{ij}=2\bar{\psi}\left(  x^{k}\right)  \bar{g}%
_{ij}~,~\bar{\psi}\left(  x^{k}\right)  =\psi\left(  x^{k}\right)  +\left(
\ln\Omega\right)  _{,k}X^{k}\text{.}%
\end{equation}
Consequently, conformally related spaces share the conformal symmetries
(CKVs).\ That is, the conformal structure remain invariant under conformal transformations.

A $n-$dimensional space, for $n\geq3$, which admits $\frac{\left(  n+1\right)
\left(  n+2\right)  }{2}$ CKVs is a maximally symmetric space and it is
conformally flat. If a space is conformally flat, then there exist a
coordinate system such that $\bar{g}_{ij}=\Omega\left(  x^{k}\right)  ^{2}%
\eta_{ij}$, where $\eta_{ij}$ is the diagonal flat space.

A main characteristic for the conformally flat spaces is that for $n=3$, the
Cotton-York tensor defined as \cite{sym4}
\begin{equation}
C_{ijk}=R_{ij;k}-R_{kj;i}+\frac{1}{4}\left(  R_{;j}g_{ik}-R_{;k}g_{ij}\right)
, \label{ss.15}%
\end{equation}
is always zero, while for $n>4$, the Weyl tensor is zero. The definition for
the Weyl tensor is as follows \cite{sym4}
\begin{equation}
C_{ijkl}=R_{ijkl}+\frac{2}{n-2}\left(  \left(  R_{i[l}g_{k]j}+R_{j[k}%
g_{l]i}\right)  +\frac{1}{\left(  n-1\right)  }Rg_{i[k}g_{l]j}\right)  .
\end{equation}
Last but not least, all two-dimensional spacetimes are conformally flat and
admits infinity number of CKVs \cite{inf1}.

CKVs are important because they can be used to identify the geometric
characteristics of a spacetime as also are related with the existence of
conservation laws for the geodesic equations. Specifically, for every KV/HKV
there correspond a conservation law for the time-like geodesic equations,
while proper CKVs are related with conservation laws for the null geodesics
\cite{ka1,ka2,ka3,kas1}.

\subsection{Conformally related Lagrangians}

Let us assume the Action Integral $S$ given by the expression%
\begin{equation}
S=\int L\left(  x^{k},\frac{dx^{k}}{ds}\right)  ds~,~ \label{ss.16}%
\end{equation}
where $L\left(  x^{k},\frac{dx^{k}}{ds}\right)  $ describes the geodesic
Lagrangian for the metric tensor $g_{ij}$, defined as
\begin{equation}
L\left(  x^{k},\frac{dx^{k}}{ds}\right)  =\frac{1}{2}g_{ij}\frac{dx^{i}}%
{ds}\frac{dx^{j}}{ds}. \label{ss.16a}%
\end{equation}

Variation of the Action Integral\ (\ref{ss.16}) gives the equations of motion,
i.e. the Euler-Lagrange equations,%
\begin{equation}
\frac{d^{2}x^{i}}{ds^{2}}+\Gamma_{jk}^{i}\left(  x^{k}\right)  \frac{dx^{j}%
}{ds}\frac{dx^{k}}{ds}=0, \label{ss.17}%
\end{equation}
where $\Gamma_{jk}^{i}\left(  x^{k}\right)  $ remarks for the the Levi-Civita
connection of the metric tensor $g_{ij}$.

For the conformal related metric $\bar{g}_{ij}=\Omega^{2}\left(  x^{k}\right)
g_{ij}$, the Lagrangian function which describes the geodesic equations is%
\begin{equation}
\bar{L}\left(  x^{k},\frac{dx^{k}}{ds}\right)  =\frac{1}{2}\Omega^{2}\left(
x^{k}\right)  g_{ij}\frac{dx^{i}}{ds}\frac{dx^{j}}{ds},
\end{equation}
where now the geodesic equations are expressed as%
\begin{equation}
\frac{d^{2}x^{i}}{ds^{2}}+\left(  \Gamma_{jk}^{i}-\left(  \ln\Omega\right)
^{,i}g_{jk}\right)  \frac{dx^{j}}{ds}\frac{dx^{k}}{ds}=0, \label{ss.18}%
\end{equation}
in which $\bar{\Gamma}_{jk}^{i}=\Gamma_{jk}^{i}-\left(  \ln\Omega\right)
^{,i}g_{jk}$ is the Levi-Civita connection for the conformally related metric
$\bar{g}_{ij}$.

It is straightforward to conclude that the geodesic equations are invariant
under a conformal transformation if and only if the Hamiltonian for the
geodesic equations is zero, that is, $\frac{1}{2}g_{ij}\frac{dx^{i}}{ds}%
\frac{dx^{j}}{ds}=0~$\cite{ka4}. Hence, the geodesic equations for Lagrangian
(\ref{ss.16a}) is invariant under conformal transformations if and only if it
is describes null geodesics.

Before we proceed with the main analysis of this study, we should remark that
for a conformally flat metric $\bar{g}_{ij}$, there exist always a coordinate
system where the null geodesic equations are expressed as terms of the linear
system%
\begin{equation}
\frac{d^{2}x^{i}}{ds^{2}}=0~,~\eta_{ij}\frac{dx^{i}}{ds}\frac{dx^{j}}{ds}=0.
\end{equation}

That interesting invariant property for the null geodesics of conformally flat
spaces we employ later in this work to study the solution space in General Relativity.

\section{The Schwarzschild spacetime}

\label{sec3}

We begin our exploration by considering the static spherically symmetric
spacetime, described by the line element%
\begin{equation}
ds^{2}=-a^{2}\left(  r\right)  dt^{2}+n^{2}\left(  r\right)  dr^{2}%
+b^{2}\left(  r\right)  \left(  d\theta^{2}+\sin^{2}\theta~d\phi^{2}\right)  .
\label{ss.01}%
\end{equation}
Only two of the three scale factors $a\left(  r\right)  $, $b\left(  r\right)
$ and $n\left(  r\right)  $ are essential, and they are determined by the
solution of the field equations. Therefore, choosing a functional form for one
of the scale factors is equivalent to selecting a coordinate system.

Within the framework of General Relativity and in vacuum, there exist a unique
analytic solution for the line element (\ref{ss.01}), derived in 1916 by Karl
Schwarzschild \cite{sch1}. \ 

In the coordinate system where $b\left(  r\right)  =r$, Schwarzschild's
solution reads
\begin{equation}
ds^{2}=-\left(  1-\frac{r_{s}}{r}\right)  dt^{2}+\left(  1-\frac{r_{s}}%
{r}\right)  ^{-1}dr^{2}+r^{2}\left(  d\theta^{2}+\sin^{2}\theta~d\phi
^{2}\right)  . \label{ss.02}%
\end{equation}
This spacetime black hole solution, where $r_{s}$ marks as the Schwarzschild radius.

Einstein's field equations for the metric tensor (\ref{ss.01}) follows from
the variation of the point-like Lagrangian function%
\begin{equation}
L\left(  n,a,a^{\prime},b,b^{\prime}\right)  =\frac{1}{2n}\left(  8ba^{\prime
}b^{\prime}+4ab^{\prime2}\right)  +2na, \label{ss.03}%
\end{equation}
where now prime denotes derivative with respect the radius parameter, that is,
$a^{\prime}=\frac{da}{dr}$.

The Euler-Lagrange equations of Lagrangian (\ref{ss.03}) leads to the
gravitational field equations
\begin{equation}
\frac{1}{2n^{2}}\left(  8ba^{\prime}b^{\prime}+4ab^{\prime2}\right)  -2a=0,
\label{ss.04}%
\end{equation}%
\begin{equation}
a^{\prime\prime}+\frac{1}{b}a^{\prime}b^{\prime}+\frac{n^{2}}{2}\frac{a}%
{b^{2}}-\frac{1}{2}\frac{a}{b^{2}}b^{\prime2}-\frac{1}{n}a^{\prime}n^{\prime
}=0, \label{ss.05}%
\end{equation}%
\begin{equation}
b^{\prime\prime}+\frac{1}{2b}b^{\prime2}-\frac{1}{a}b^{\prime}n^{\prime}%
-\frac{1}{2}\frac{n^{2}}{b}=0. \label{ss.06}%
\end{equation}

The field equations (\ref{ss.04}), (\ref{ss.05}), (\ref{ss.06}) form a
singular Hamiltonian system with equation (\ref{ss.04}) to be the Hamiltonian constraint.

We introduce the momentum $p_{a}=\frac{\partial L}{\partial a^{\prime}}$,
$p_{b}=\frac{\partial L}{\partial b^{\prime}}$, and in the Hamiltonian
formalism the field equations (\ref{ss.04}), (\ref{ss.05}), (\ref{ss.06})
become%
\begin{equation}
n\left(  \frac{p_{a}p_{b}}{4b}-\frac{a}{8b^{2}}p_{a}^{2}-2a\right)  =0~,
\label{ss.07}%
\end{equation}%
\begin{equation}
\frac{1}{n}a^{\prime}=\frac{ap_{a}-bp_{b}}{4b^{2}}~,~\frac{1}{n}b^{\prime
}=\frac{p_{a}}{4b}~, \label{ss.08}%
\end{equation}%
\begin{equation}
\frac{1}{n}p_{a}^{\prime}=2+\frac{p_{a}^{2}}{b^{2}}~,~\frac{1}{n}%
p_{b}^{^{\prime}}=\frac{1}{4b^{2}}\left(  p_{a}p_{b}-\frac{p_{a}^{2}}%
{2}\right)  ~. \label{ss.09}%
\end{equation}

These equations are of the same form as those of a Hamiltonian system, which
describes the motion of two particles with varying mass under the action of a
conservative force. Here, the scale factors play the role of the particles,
while the spatial curvature term provides the interaction term.

We employ Eisenhart's approach \cite{el} to write the field equations
(\ref{ss.07}), (\ref{ss.08}), (\ref{ss.09}) in the equivalent form of a
Hamiltonian system which describes geodesic equations. Indeed, we introduce
the new scalar $\psi$, and the momentum $p_{\psi}$, where the new Hamiltonian
function is
\begin{equation}
\mathcal{H}=n\left(  \frac{p_{a}p_{b}}{4b}-\frac{a}{8b^{2}}p_{a}^{2}%
-2ap_{\psi}^{2}\right)  . \label{ss.10}%
\end{equation}
Consequently, the equations of motion in terms of the momentum are written as
follows
\begin{equation}
n\left(  \frac{p_{a}p_{b}}{4b}-\frac{a}{8b^{2}}p_{a}^{2}-2ap_{\psi}%
^{2}\right)  =0 \label{ss.11}%
\end{equation}%
\begin{equation}
\frac{1}{n}a^{\prime}=\frac{ap_{a}-bp_{b}}{4b^{2}}~,~\frac{1}{n}b^{\prime
}=\frac{p_{a}}{4b}~,~\frac{1}{n}\psi^{\prime}=4ap_{\psi}~, \label{ss.12}%
\end{equation}%
\begin{equation}
\frac{1}{n}p_{a}^{\prime}=2p_{z}^{2}+\frac{p_{a}^{2}}{b^{2}}~,~\frac{1}%
{n}p_{b}^{^{\prime}}=\frac{1}{4b^{2}}\left(  p_{a}p_{b}-\frac{p_{a}^{2}}%
{2}\right)  ~,~p_{\psi}^{\prime}=0~. \label{ss.13}%
\end{equation}

From the latter expression, that is, equation (\ref{ss.13}), it follows that
the momentum $p_{\psi}$ is conserved. That is, $p_{\psi}$ represents a second
conservation law for the geodesic equations. Nevertheless, in order to recover
the original gravitational system (\ref{ss.07}), (\ref{ss.08}), (\ref{ss.09}),
the following constraint should be applied, $p_{\psi}=1$.

As we shall see in the following lines, the introduction of the scalar field
$\psi$ leads to the introduction of new conservation laws, and new dynamical
properties. These conservation laws are not lost when conservation law
$p_{\psi}=1$ is applied in the system, but they become nonlocal, that is,
hidden symmetries.

The Hamiltonian function (\ref{ss.10}) with the constraint (\ref{ss.11})
describes the null geodesic equations for the three-dimensional space with
line element%
\begin{equation}
ds^{2}=n\left(  8b~da~db+4a~db^{2}-\frac{d\psi^{2}}{2a}\right)  .
\label{ss.14}%
\end{equation}
We will refer to the latter space as the extended minisuperspace.

The null geodesics are invariant under conformation, which is why parameter
$n$ plays no role in the dynamics.

Thus, for the line element (\ref{ss.14}) we calculate that all the components
of the Cotton-York tensor (\ref{ss.15}) are zero; that is $C_{ijk}=0$. This
property states that the three-dimensional space (\ref{ss.14}) is conformally
flat, that is, there exist a coordinate transformation $\left\{
a,b,\psi\right\}  \rightarrow\left\{  x,y,z\right\}  $, where the line element
is of the form $ds^{2}=n\Omega^{2}\left(  x,y,z\right)  \left(  \alpha
_{1}dx^{2}+\alpha_{2}dy^{2}+\alpha_{3}dz^{2}\right)  $, where $\alpha
_{1},~\alpha_{2}$ and $\alpha_{3}$ are constants. Function $\Omega\left(
x,y,z\right)  $ is known as the conformal factor. In the new coordinates
$\left\{  x,y,z\right\}  $, the equations of motion (\ref{ss.11}),
(\ref{ss.12}), (\ref{ss.13}) are linear. Recall that the unique linear
geodesic equations are those of the free particle in the flat space.

An equivalent way to verify this property is to calculate the number of
conservation laws for the null geodesics. It is known that Conformal Killing
Vectors (CKVs) generate conservation laws for null geodesics. Hence, the
conformal condition for the line element (\ref{ss.14}) leads to the derivation
of ten CKVs, which is the maximum number of CKVs for a three dimensional
space; that is, space (\ref{ss.14}) is conformally flat.

We introduce the new variable $A$, with the transformation rule $a=\sqrt
{\frac{A}{b}}$, then the line element (\ref{ss.14}) reads%
\begin{equation}
ds^{2}=\frac{1}{n}\left(  \frac{b}{A}\right)  ^{\frac{1}{2}}\left(
8~dAdb-d\psi^{2}\right)  .
\end{equation}

By introducing the diagonal coordinates $A=\frac{x+y}{2\sqrt{2}}%
~,~b=\frac{x-y}{2\sqrt{2}}$, it follows%
\begin{equation}
ds^{2}=\frac{1}{n}\left(  \frac{x-y}{x+y}\right)  ^{\frac{1}{2}}\left(
dx^{2}-dy^{2}-d\psi^{2}\right)  ,
\end{equation}

Coordinates $\left\{  x,y,\psi\right\}  $ are the canonical coordinates for
the Hamiltonian system (\ref{ss.10}). In the coordinate system$\left\{
x,y,\psi\right\}  $, the field equations (\ref{ss.11}), (\ref{ss.12}),
(\ref{ss.13}) are written in the following linearized form%
\begin{equation}
\frac{1}{\tilde{n}}x^{\prime}=p_{x}~,~\frac{1}{\tilde{n}}y^{\prime}%
=p_{y}~,~\frac{1}{\tilde{n}}\psi^{\prime}=p_{\psi},
\end{equation}%
\begin{equation}
p_{x}^{\prime}=0~,~p_{y}^{\prime}=0~,~p_{\psi}^{\prime}=0,
\end{equation}
with constraints%
\begin{equation}
x^{\prime2}-y^{\prime2}-\psi^{\prime2}=0,~p_{\psi}=1, \label{con.0}%
\end{equation}
and $\tilde{n}=n\left(  \frac{x-y}{x+y}\right)  ^{-\frac{1}{2}}$.

Without loss of generality we can select $\tilde{n}=1$ and the latter
dynamical system takes the form of the free particle in a three dimensional
flat space, that is,%
\begin{equation}
x^{\prime\prime}=0~,~y^{\prime\prime}=0~,~\psi^{\prime\prime}=0, \label{ss.19}%
\end{equation}
with constraints (\ref{con.0}).

We have demonstrated that the solution space for the Einstein field equations
for this problem corresponds to that of the three-dimensional free particle in
a flat space. It is important to note that we are referring to the dynamics of
the scale factors driven by the gravitational theory, and not on the test
particles of the physical space. While the transformation applied to linearize
the field equations is not unique, the uniqueness lies in the solution of the
field equations itself.

\section{De Sitter-Schwarzschild spacetime}

\label{sec4}

The introduction of the cosmological constant $\Lambda$ within the framework
of the static spherical symmetric spacetime (\ref{ss.01}) leads to the de
Sitter-Schwarzschild metric with line element \cite{sch2}
\begin{equation}
ds^{2}=-\left(  1-\frac{r_{s}}{r}-\frac{\Lambda}{3}r^{2}\right)
dt^{2}+\left(  1-\frac{r_{s}}{r}-\frac{\Lambda}{3}r^{2}\right)  ^{-1}%
dr^{2}+r^{2}\left(  d\theta^{2}+\sin^{2}\theta~d\phi^{2}\right)  .
\label{ss.20}%
\end{equation}
We observe that the Schwarzschild spacetime (\ref{ss.02}) is recovered in the
limit where the cosmological constant vanishes.

The point-like Lagrangian which describes the evolution of the scale factors,
leading to the analytic solution (\ref{ss.20}), is as follows
\begin{equation}
L^{\Lambda}\left(  n,a,a^{\prime},b,b^{\prime}\right)  =\frac{1}{2n}\left(
8ba^{\prime}b^{\prime}+4ab^{\prime2}\right)  +2na\left(  1+\Lambda
b^{2}\right)  . \label{ss.21}%
\end{equation}

We employ the same procedure as before. The equivalent geodesic Hamiltonian,
which describes the field equations for the de Sitter-Schwarzschild geometry,
is
\begin{equation}
\mathcal{H}^{\Lambda}=n\left(  \frac{p_{a}p_{b}}{4b}-\frac{a}{8b^{2}}p_{a}%
^{2}-2a\left(  1+\Lambda b^{2}\right)  p_{\psi}^{2}\right)  , \label{ss.22}%
\end{equation}
with constraints $\mathcal{H}^{\Lambda}=0$ and $p_{\psi}=1$.

Furthermore, the line element for the corresponding extended minisuperspace
is
\begin{equation}
ds^{\Lambda~2}=\frac{1}{n}\left(  8b~da~db+4a~db^{2}-\frac{d\psi^{2}%
}{2a\left(  1+\Lambda b^{2}\right)  }\right)  . \label{ss.23}%
\end{equation}

For the three-dimensional space (\ref{ss.23}) the Cotton-York tensor
(\ref{ss.15}) has zero components, that is, space (\ref{ss.23}) has the
maximum conformal algebra and it is conformally flat.

We consider the same change of variables as before $a=\sqrt{\frac{A}{b}}$,
such that the line element (\ref{ss.23}) is expressed as follows
\begin{equation}
ds^{\Lambda~2}=\frac{1}{\left(  1+\Lambda b^{2}\right)  n}\left(  \frac{b}%
{A}\right)  ^{\frac{1}{2}}\left(  8\left(  1+\Lambda b^{2}\right)
dAdb-d\psi^{2}\right)  . \label{ss.24}%
\end{equation}

Under the second change of variables $dB=\int\left(  1+\Lambda b^{2}\right)
db,$ it follows $ds^{\Lambda~2}=\frac{1}{\hat{n}}\left(  8dAdB-d\psi
^{2}\right)  ~$where $\hat{n}=\left(  1+\Lambda b^{2}\right)  n\left(
\frac{b}{A}\right)  ^{\frac{1}{2}}$.

Finally in the diagonal variables $A=\frac{X+Y}{2\sqrt{2}}$ and $B=\frac
{X-Y}{2\sqrt{2}}$, the extended minisuperspace is written in the canonical
form of a conformally flat space, that is,%
\begin{equation}
ds^{\Lambda~2}=\frac{1}{2\hat{n}}\left(  dX^{2}-dY^{2}-d\psi^{2}\right)  .
\label{ss.25}%
\end{equation}
Consequently the gravitational field equations are written in the equivalent
form of the free particle in a three-dimensional flat space, i.e.,
\begin{equation}
X^{\prime\prime}=0~,~Y^{\prime\prime}=0~,~\psi^{\prime\prime}=0, \label{ss.26}%
\end{equation}
with constraint equation%
\begin{equation}
X^{\prime2}-Y^{\prime2}-\psi^{\prime2}=0~,~p_{\psi}=1. \label{ss.27}%
\end{equation}

We remark that the field equations for the Schwarzschild black hole, whether
in a Minkowski or a de Sitter background, share a common solution space, which
is that of the null geodesic equations in a conformally flat extended
minisuperspace. Consequently, there exists a one-to-one transformation that
relates the two solutions. It's important to note that this transformation
does not relate the physical space but rather the space of solutions for the
scale factors of spacetime (\ref{ss.01}).

At this point we want to mention that this is not the unique approach to
extract the de Sitter-Schwarzschild from the Schwarzschild geometry. Another
geometric construction approach can be found in \cite{jb1,jb2,jb3,jb4}.

We now proceed with our investigation into the solution space when an
electromagnetic fluid is introduced into the physical space.

\section{The Reissner-Nordstr\"{o}m black hole}

\label{sec5}

The analytic solution of Einstein's General Relativity for a static spherical
symmetric spacetime (\ref{ss.01}) with charge is the Reissner-Nordstr\"{o}m
black hole \cite{sch3,sch4}%
\begin{equation}
ds^{2}=-\left(  1-\frac{r_{s}}{r}+\frac{r_{Q}^{2}}{r^{2}}\right)
dt^{2}+\left(  1-\frac{r_{s}}{r}+\frac{r_{Q}^{2}}{r^{2}}\right)  ^{-1}%
dr^{2}+r^{2}\left(  d\theta^{2}+\sin^{2}\theta~d\phi^{2}\right)  ,
\label{ss.30}%
\end{equation}
in which $r_{Q}$ is the characteristic length scale related to the charge.

The field equations are described by the point-like Lagrangian \cite{dim1}
\begin{equation}
L^{RN}\left(  n,a,a^{\prime},b,b^{\prime},\zeta,\zeta^{\prime}\right)
=\frac{1}{2n}\left(  8ba^{\prime}b^{\prime}+4ab^{\prime2}+4\frac{b^{2}}%
{a}\zeta^{\prime2}\right)  +2na, \label{ss.31}%
\end{equation}
where $\zeta\left(  r\right)  $ is the potential of the electromagnetic tensor.

For this dynamical system, the corresponding geodesic equivalent Hamiltonian
is
\begin{equation}
\mathcal{H}^{RN}=n\left(  \frac{p_{a}p_{b}}{4b}-\frac{a}{8b^{2}}p_{a}%
^{2}+\frac{a}{8b^{2}}p_{\zeta}^{2}-2ap_{\psi}^{2}\right)  \label{ss.32}%
\end{equation}
with constraints $\mathcal{H}^{\Lambda}=0$ and $p_{\psi}=1$. The extended
minisuperspace is defined as
\begin{equation}
ds^{RN~2}=\frac{1}{n}\left(  8b~da~db+4a~db^{2}+4\frac{b^{2}}{a}d\zeta
^{2}-\frac{d\psi^{2}}{2a}\right)  . \label{ss.33}%
\end{equation}

For the latter element the Weyl tensor is calculated to be always zero.
Consequently, the extended minisuperspace (\ref{ss.33}) is conformally flat.

We introduce the change of variables $a=\sqrt{\frac{A}{b}+\frac{z^{2}}{b^{2}}%
}$,~$~\zeta=\frac{z}{b^{2}}$. Hence, the extended minisuperspace (\ref{ss.33})
is expressed%
\begin{equation}
ds^{RN~2}=\frac{1}{n}\frac{b}{\sqrt{bA+z^{2}}}\left(  4dA~db+4dz^{2}-d\psi
^{2}\right)  , \label{ss.34}%
\end{equation}
where easily it can be written in the diagonal form%
\begin{equation}
ds^{RN~2}=\frac{1}{\check{n}}\left(  dU^{2}-dV^{2}-dZ^{2}-d\psi^{2}\right)  ,
\label{ss.35}%
\end{equation}
and the field equations take the linear form%
\begin{equation}
U^{\prime\prime}=0~,\ V^{\prime\prime}=0~,~Z^{\prime\prime}=0~,~\psi
^{\prime\prime}=0~, \label{ss.36}%
\end{equation}
with constraints%
\begin{equation}
U^{\prime2}-V^{\prime2}-Z^{\prime2}-\psi^{\prime2}=0~,~p_{\psi}=1.
\label{ss.37}%
\end{equation}

We observe that the solution space for the field equations of the
Reissner-Nordstr\"{o}m black hole consists once again of the equations of
motion for a free particle in a flat geometry. This property is similar to the
solution space for the field equations of the Schwarzschild and de
Sitter-Schwarzschild spacetimes. However, the dimension of the solution space
is higher due to the additional degrees of freedom related to the charge.

\section{Bianchi II vacuum spacetime}

\label{sec6}

Let us proceed our discussion with the consideration of cosmological
spacetimes. We consider the locally rotational Bianchi II geometry with the
line element%
\begin{equation}
ds^{2}=-N^{2}\left(  t\right)  dt^{2}+a\left(  t\right)  ^{2}\left(  dr-\theta
d\phi\right)  ^{2}+b\left(  t\right)  ^{2}\left(  d\theta^{2}+d\phi
^{2}\right)  . \label{ss.38}%
\end{equation}
For this gravitational model the point-like Lagrangian which reproduces the
field equations is defined as%
\begin{equation}
L^{II}\left(  N,a,\dot{a},b,\dot{b}\right)  =\frac{1}{N}\left(  2b\dot{a}%
\dot{b}+a\dot{b}^{2}\right)  +N\frac{a^{3}}{b^{2}}, \label{ss.39}%
\end{equation}
where a dot denotes derivative with respect to the time parameter, i.e.
$\dot{a}=\frac{da}{dt}$. \ The vacuum solution derived before in \cite{va1}.

The Hamiltonian function for the geodesic description of the field equations
is
\begin{equation}
\mathcal{H}^{II}=N\left(  \frac{p_{a}p_{b}}{2b}-\frac{a}{4b^{2}}p_{a}%
^{2}-\frac{a^{3}}{b^{2}}p_{\psi}^{2}\right)  , \label{ss.40}%
\end{equation}
with the constraints $\mathcal{H}^{II}=0$ , $p_{\psi}=1$.

Therefore the extended minisuperspace has the following line element%
\begin{equation}
ds^{II~2}=\frac{1}{N}\left(  4b~da~db+2a~db^{2}-\frac{b^{2}}{a^{3}}d\psi
^{2}\right)  , \label{ss.41}%
\end{equation}
where easily it follows that the line element (\ref{ss.41}) is conformally flat.

In the terms of the new dynamical variables $a\rightarrow\left(  AB\right)
^{\frac{1}{4}}$ and $b\rightarrow B^{-\frac{1}{2}}$, the extended
minisuperspace becomes%
\begin{equation}
ds^{II~2}=\frac{1}{A^{\frac{3}{4}}B^{\frac{7}{4}}N}\left(  \frac{1}%
{2}dA~dB+d\psi^{2}\right)  . \label{ss.42}%
\end{equation}

Therefore, the field equations can be written in the equivalent form of the
linearized system%
\begin{equation}
\ddot{x}=0~,~\ddot{y}=0~,~\ddot{\psi}=0~, \label{ss.43}%
\end{equation}
in which we have introduced the second change of variables $A=\sqrt{2}\left(
x+y\right)  ~,~B=\sqrt{2}\left(  x-y\right)  $. Finally, the following
constraints hold true $\dot{x}^{2}-\dot{y}^{2}-\dot{\psi}^{2}=0$, $p_{\psi}=1$.

\section{The Lie algebra $D\otimes_{s}T_{2}$}

\label{sec7}

In this Section, we investigate the common geometric property of the field
equations discussed in the previous sections. This property leads to the
unification and linearization of Einstein's field equations, resulting in the
trivial derivation of the analytic solutions.

The field equations generated by the point-like Lagrangian (\ref{ss.03}) of
the Schwarzschild spacetime are invariant under the point transformations with
generators, the vector fields.%
\[
X^{1}=\frac{1}{ab}\partial_{a}~,~X^{2}=-a\partial_{a}+b\partial_{b}~,
\]%
\[
X^{3}=\left(  -\frac{a}{2b}\partial_{a}+\partial_{b}\right)  .
\]
For these three vector fields we calculate the the commutators
\[
\left[  X^{1},X^{2}\right]  =-X^{1}~,~\left[  X^{1},X^{3}\right]  =0~,~\left[
X^{2},X^{3}\right]  =-X^{2}\,.
\]
Therefore, the vector fields $\left\{  X^{1},X^{2},X^{3}\right\}  $ form the
Lie algebra $A_{3,3}$ in the Patera et al. classification scheme
\cite{patera}. It is a solvable Lie algebra commonly known as the semidirect
product of dilations and translations in the plane, i.e., $D\otimes_{s}%
T_{2}\equiv A_{1}\otimes_{s}2A_{1}$.

In the appearance of the cosmological constant in the physical background
space, the field equations for the de Sitter-Schwarzschild are invariant under
the point transformations with generators%
\[
X_{\Lambda}^{1}=\frac{1}{ab}\partial_{a}~,~X_{\Lambda}^{2}=\frac{1}{1+\Lambda
b^{2}}\left(  -a\left(  1+\frac{2}{3}\Lambda b^{2}\right)  \partial
_{a}+b\left(  1+\frac{\Lambda}{3}b^{2}\right)  \partial_{b}\right)  ~,
\]%
\[
X_{\Lambda}^{3}=\frac{1}{\left(  1+\Lambda b^{2}\right)  }\left(  -\frac
{a}{2b}\partial_{a}+\partial_{b}\right)  .
\]
The vector fields $\left\{  X_{\Lambda}^{1},X_{\Lambda}^{2},X_{\Lambda}%
^{3}\right\}  $ have the same commutator rules with that for $\Lambda=0$,
which means that they form the $D\otimes_{s}T_{2}$ Lie algebra, expressed in a
different representation.

As far as the dynamical system described by the point-like Lagrangian
(\ref{ss.31}) is concerned; that is, the field equations for the
Reissner-Nordstr\"{o}m spacetime, they are invariant under the point
transformations with generators the following vector fields%
\[
X_{RN}^{1}=\frac{1}{ab}\partial_{a}~,~X_{RN}^{2}=-a\partial_{a}+b\partial
_{b}-z\partial_{\zeta}~,~
\]%
\[
X_{RN}^{3}=-\left(  \frac{a}{2b}+\frac{z^{2}}{ab}\right)  \partial
_{a}+\partial_{b}-\frac{\zeta}{b}\partial_{\zeta}~,
\]%
\[
X_{RN}^{4}=-a\zeta\partial_{a}+b\zeta\partial_{b}+\left(  \frac{a^{2}}%
{4}-\frac{z^{2}}{2}\right)  \partial_{\zeta},
\]%
\[
X_{RN}^{5}=\frac{2\zeta}{ab}\partial_{a}+\frac{1}{b}\partial_{\zeta}%
~,~X_{RN}^{6}=\partial_{\zeta}.
\]
The vector fields $\left\{  X_{RN}^{1},X_{RN}^{2},X_{RN\ }^{3}\right\}  $,
form the $D\otimes_{s}T_{2}$ subalgebra.

Finally, the cosmological field equations for the Bianchi II geometry
described by the point-like Lagrangian function (\ref{ss.39}) are also
invariant under the family of point transformations with generators the vector
fields%
\[
X_{II}^{1}=\frac{1}{a^{3}b^{2}}\partial_{a}~,~X_{II}^{2}=b\partial_{b}~,
\]%
\[
X_{II}^{3}=-\frac{ab^{2}}{2}\partial_{a}+b^{3}\partial_{b}.
\]
The latter vector fields form again the $D\otimes_{s}T_{2}$ algebra.

We conclude that the common feature of these four-different models that we
proved that they are linearisable is the existence of the Lie symmetry vectors
which form the three-dimensional $A_{3,3}$ or equivalent, the $D\otimes
_{s}T_{2}$ Lie algebra. However, the natural question which arise is what is
the origin of the $D\otimes_{s}T_{2}$ Lie algebra, and how it is related with
the linearization process.

Consider for instance the maximum symmetric linear system (\ref{ss.19}). The
dynamical system admits ten symmetry vectors, due to the constraint equations.
However, the application of the constraint $p_{\psi}=1$, in order to determine
the original system, indicates that only three of the ten symmetries remain
points, while the rest six vector fields become nonlocal. The three symmetries
which survive are those which form the $D\otimes_{s}T_{2}$ Lie algebra.

In the case of the higher-dimensional linear system (\ref{ss.36}), the
admitted Lie symmetries are fifteen, where only the six symmetries remain
point symmetries when the constraint $p_{\psi}=1$ is applied.

Therefore, when a gravitational system is invariant under point
transformations with generators the elements of the $D\otimes_{s}T_{2}$ Lie
algebra, we have a strong indication that this given dynamical system can be
linearized, and the closed-form solution of the field equations can be written
in analytic form.

\section{Conclusions}

\label{sec8}

In this piece of work, we delved into the solution space of Einstein's General
Relativity for several well-known spacetimes. Specifically, we focused on
investigating the solution space for the gravitational field equations
governing the following spacetimes, Schwarzschild vacuum solution, with or
without the cosmological constant term, the Reissner-Nordstr\"{o}m black hole
with a charge and the vacuum solution for the locally rotational Bianchi II cosmology.

For the aforementioned geometries, the Einstein's field equations are
invariant under the action of point transformations which form the same Lie
algebra, the $D\otimes_{s}T_{2}$ Lie algebra, also known as the $A_{3,3}~$Lie
algebra. This specific Lie algebra originates from a higher-dimensional
equivalent dynamical system that describes geodesic equations in the solution
space for the field equations.

For each gravitational model in our study, the equivalent higher-dimensional
dynamical system is found to correspond to the null geodesic equations of a
conformally flat geometry. Hence, we were able to determine coordinate
transformations where the field equations for each gravitational model can be
expressed in terms of the equations of motion for the Newtonian free particle
in three- (or four-) dimensional space. The families of these coordinate
transformations are those that relate the different representations of the
admitted symmetries for the $D\otimes_{s}T_{2}$ Lie algebra.

The static spherically symmetric spacetime considered previously is directly
related to the Kantowski-Sachs geometry and the locally rotational Bianchi III
geometry. Thus, the results of this analysis are valid not only for the
Einstein's field equations governing the static spherically symmetric
spacetime but also for these two other spacetimes.

The $D\otimes_{s}T_{2}$ Lie algebra for some of the above gravitational models
has been determined before in \cite{sym9,dim1,dim2} by applying the method of
variational symmetries in the original minisuperspace Lagrangian. For an
extended minisuperspace, and specifically for the Eisenhart-Duval lift and in
the case of the Lorentzian lift, the $D\otimes_{s}T_{2}$ Lie algebra has been
determined before for the static spherically symmetric spacetimes in
\cite{kas1,jb1}. However, the definition of the Eisenhart lift is not unique
and the admitted symmetries for the extended minisuperspace depend on the
lift. But the $D\otimes_{s}T_{2}$ Lie algebra is preserved by the lift. In my
consideration, I followed a different lift, and I applied the Riemannian lift. 

The focus of this study is to identify the property the field equations for
these models can be linearized, by using simple geometric techniques. Indeed,
the geometric linearization is equivalent with the existence and the
construction of conservation laws. What is more, is the common property for
all these systems, the existence of the $D\otimes_{s}T_{2}$ Lie algebra. The
origin of the $D\otimes_{s}T_{2}$ Lie algebra follows from the Conformal
symmetries of the extended minisuperspace, where when we apply the new
conservation law to eliminate the lift, only the elements of the $D\otimes
_{s}T_{2}$ \ Lie algebra survive as local symmetries. At this point, it is
important to mention that the application of the Lorentzian lift in the above
systems, as in the studies in \cite{kas1,jb1}, lead to extended minisuperspace
which admit additional symmetries, but they are not conformally
flat.\ Consequently, the field equations can not be\ geometric linearized via
the Lorentzian lift. 

We conclude that the common solution space for these gravitational models is
the solution to the linear equations of the Newtonian free particle. This
geometric approach opens new directions for deriving analytic solutions in
gravitational physics. It extends the application of the harmonic maps
\cite{hm1,hm2,hm3} in gravitational physics. Furthermore, this method can be
applied to modified theories of gravity and dark energy cosmological models.
In future work, we plan to further investigate these considerations.

\textbf{Data Availability Statements:} Data sharing is not applicable to this
article as no datasets were generated or analyzed during the current study.

\begin{acknowledgments}
AP thanks the support of VRIDT through Resoluci\'{o}n VRIDT No. 096/2022 and
Resoluci\'{o}n VRIDT No. 098/2022.
\end{acknowledgments}

\end{document}